\documentclass{aa}
\usepackage{graphics}

\newcommand{\h}{$^{\rm h}$}
\newcommand{\m}{$^{\rm m}$}
\newcommand{\s}{$^{\rm s}$}

\newcommand{\ha}{\rm H$\alpha$}
\newcommand{\hbeta}{\rm H$\beta$}
\newcommand{\HII}{\ion{H}{ii}}
\newcommand{\hnii}{{\rm H}$\alpha+[$\ion{N}{ii}$]$}
\newcommand{\nii}{$[$\ion{N}{ii}$]$}
\newcommand{\sii}{$[$\ion{S}{ii}$]$}

\newcommand{\oiii}{$[$\ion{O}{iii}$]$}

\newcommand{\flux}{$10^{-17}$ erg s$^{-1}$ cm$^{-2}$ arcsec$^{-2}$}
\newcommand{\dens}{\rm cm$^{-3}$}
\newcommand{\vel}{\rm km s$^{-1}$}
\newcommand{\sulfur}{[S~{\sc ii}]}
\newcommand{\nitrogen}{[N~{\sc ii}]}
\newcommand{\oxygen}{[O~{\sc iii}]}
\newcommand{\siirat}{$[$\ion{S}{ii}$]\lambda\ 6716/6731$} 
\begin{document}

%   \thesaurus{08     % A&A Section 8: Diffuse matter in space
%              (09.07.01;  % ISM : general,
%		09.19.2;
%		09.09.1)} % Superona remnants
%
\title{New optical filamentary structures in Pegasus}

\author{P. Boumis\inst{1}
\and  F. Mavromatakis\inst{1}
\and E. V. Paleologou\inst{1}
\and W. Becker\inst{2}
}
\offprints{P. Boumis,~~\email{ptb@physics.uoc.gr}}
\authorrunning{P. Boumis et al.}
\titlerunning{New optical filamentary structures in Pegasus}
\institute{
University of Crete, Physics Department, P.O. Box 2208, 710 03 Heraklion, Crete, Greece 
\and Max-Planck Institut f\"{u}r extraterrestrische Physik, Giessenbachstrasse, 85740 Garching, Germany}
\date{Received 01 July 2002 / Accepted 11 September 2002}

\abstract{Deep \ha$+$\nitrogen~CCD images have been obtained in the
area of the Pegasus Constellation. The resulting mosaic covers an
extent of $\sim$ 7\degr.5$\times$~8\degr.5 and filamentary and diffuse
emission is discovered. Several long filaments (up to $\sim$1\degr)
are found within the field, while diffuse emission is present mainly
in the central and northern areas. The filaments show variations in
the intensity along their extent suggesting inhomogeneous interstellar
clouds. Faint soft X--ray emission is also detected in the ROSAT
All--Sky Survey. It is mainly concentrated in the central areas of our
field and overlaps the optical emission. The low ionization images of
\sulfur\ of selected areas mainly show faint diffuse emission, while
in the medium ionization images of \oxygen\ diffuse and faint
filamentary structures are present. Spectrophotometric observations
have been performed on the brightest filaments and indicate emission
from photoionized or shock--heated gas. The sulfur line ratios
indicate electron densities below $\sim$600 cm$^{-3}$, while the
absolute \ha\ emission lies in the range of 1.1 -- 8.8 $\times$
\flux. The detected optical line emission could be part of a single or
multiple supernova explosions.
 }

\maketitle

\section{Introduction}
Several surveys have been made in the last decade concerning
galactic supernova remnants (SNRs -- Arendt 1989; Seward 1990; Koo \&
Heiles 1991; Saken et al. 1992;). Green (2001) published a revised
catalogue containing 231 Galactic SNRs and details for a number of
possible and probable SNRs. In an effort to deepen our knowledge on
the properties of the optically detected remnants, many imaging and
spectral observations have been performed (Fesen \& Hurford 1995,
Fesen et al. 1995, 1997; Boumis et al. 2001; Mavromatakis et al. 2001,
2002a, 2002c) while, new optical SNRs have also been discovered
(Boumis et al. 2002; Mavromatakis \& Strom 2002; Mavromatakis et
al. 2002b).

The Pegasus constellation is a region without any historical record of
SN events. The Palomar Observatory Sky Survey (POSS) plates do
not provide clear evidence of optical emission however, careful
examination of the POSS plates reveals very weak extended optical
emission. In particular, two of the filaments were detected on the
POSS plates and we considered that such structures in high galactic
latitudes would be interesting to study. The imaged area was expanded
as more emission line structures were discovered. However, the final
field size was restricted by the available telescope time. The
published radio maps do not provide evidence for excess non--thermal
emission which could be attributed to a supernova remnant.

In this paper, we report the discovery of faint optical filamentary
and diffuse X--ray emission from the Pegasus constellation. We present
an \hnii\ mosaic which covers an area of
$\sim$7\degr.5$\times$~8\degr.5, and \sii~and \oiii~images of selected
regions showing filamentary and diffuse structures. Spectrophotometric
observations of the brightest filaments were also performed and the
emission lines were measured. In Sect. 2, we present information
concerning the observations and data reduction, while the results of
the imaging and spectral observations are given in Sect. 3 and 4,
respectively. Information about the X--ray data are given in Sect. 5,
while in the last section (Sect. 6) we discuss the physical properties
of the newly detected structures.

\section{Observations}
\subsection{Imaging}

\begin{table*}
\caption{Parameters of observations}
\label{table1}
\begin{flushleft}
\begin{tabular}{lccclc}
\hline 
\hline
\multicolumn{6}{c}{IMAGING} \\
\hline
Date &
Telescope &
Detector &
Filter &
$\lambda_{\rm c}$~-- FWHM (\AA) &
Object \\
\hline
2000 Aug 25--Sep 01 & 0.3 m & Thomson$^{\rm a}$ & \ha\ $+$ \nitrogen\ & 6560 -- 75 & Mosaic \\ 
2000 Aug 25--Sep 01 & 0.3 m & Thomson & Cont red & 6096 -- 134 & Mosaic \\
2000 Sep 02--03 & 0.3 m & Thomson & \sulfur\ & 6708 -- 20 & F3--F6 \\
2000 Sep 02--03 & 0.3 m & Thomson & Cont red & 6096 -- 134 & F3--F6 \\
2000 Sep 03--04 & 0.3 m & Thomson & \oxygen\ & 5005 -- 28 & F3--F6 \\
2000 Sep 03--04 & 0.3 m & Thomson & Cont blue & 5470 -- 230 & F3--F6 \\
2001 Aug 10--11 & 1.3 m & SITe$^{\rm b}$ & \ha\ $+$ \nitrogen\ & 6570 -- 75 & F3b, F5a, F6 \\
\hline
\multicolumn{6}{c}{SPECTROSCOPY} \\
\hline
Date &
Telescope &
Detector &
Slit width ($\mu$m) &
Waveband (\AA) &
Object \\
\hline
2001 Aug 12, 20--24  & 1.3 m & SITe$^{\rm c}$ & 320 & 4750 -- 6815 & F2--F6 \\
\hline
\end{tabular}
\end{flushleft}
${\rm ^a}$ 70\arcmin$\times$70\arcmin, 4\arcsec.12 pixel$^{-1}$\\
${\rm ^b}$ 8\arcmin.5$\times$8\arcmin.5, 0\arcsec.5 pixel$^{-1}$\\
${\rm ^c}$ 7\arcmin.9$\times$20\arcmin, 0\arcsec.59 pixel$^{-1}$\\
\end{table*}

The image of the filamentary and diffuse nebulosities shown in Fig. 1
is a mosaic of 27 images taken through an \ha$+$\nitrogen~filter with
the 0.3 m Schmidt--Cassegrain (f/3.2) telescope at Skinakas
Observatory in Crete, Greece from August 25 to September 04, 2000. The
1024$\times$1024 (with 19$\times$19 $\mu$m$^{2}$ pixels) Thomson CCD
was used resulting in a scale of 4\arcsec.12 pixel$^{-1}$~and a
field of view of 70\arcmin~$\times$~70\arcmin. The brightest filaments
named F3, F4, F5 and F6 in Fig. 1 were also observed using the same
configuration with the \oxygen\ and \sulfur\ emission line
filters. One 1800 s exposure in the light of
\ha$+$\nitrogen\ and two in the light of \sulfur~and
\oxygen~(bringing their total exposure times to 3600 s) were taken for 
each field. Optical images of filaments F3, F5 and F6 were also
obtained through the \ha$+$\nitrogen~filter (Fig. 5) with the 1.3 m
(f/7.7) Ritchey--Cretien telescope at Skinakas Observatory on August
10--11, 2001. The detector was the 1024$\times$1024 (with 24$\times$24
$\mu$m$^{2}$ pixels) SITe CCD giving a field of view of
8\arcmin.5$\times$~8\arcmin.5 and an image scale of 0\arcsec.5
pixel$^{-1}$. Details of all observations are summarized in
Table~\ref{table1}. 

The image reduction was carried out using the IRAF and MIDAS packages
and their grey-scale representation using the STARLINK Kappa and
Figaro packages. The astrometric solutions were calculated for each
field using reference stars from the Hubble Space Telescope (HST)
Guide Star Catalogue (Lasker et al. 1999). All coordinates quoted in
this paper refer to epoch 2000.

\subsection{Spectroscopy}
Low dispersion long--slit spectra were obtained with the 1.3 m
telescope at Skinakas Observatory on August 12 and August 20-24,
2001. The 1300 line mm$^{-1}$~grating was used in conjunction with a
2000$\times$800 SITe CCD (15$\times$15 $\mu$m$^{2}$~pixels) resulting
in a scale of 1 \AA\ pixel$^{-1}$ and covers the range of 4750 \AA\ --
6815 \AA. The spectral resolution is $\sim$8 pixels and $\sim$11
pixels full width at half maximum (fwhm) for the red and blue
wavelengths, respectively. The slit width is 7\farcs7 and in all
cases, was oriented in the south--north direction, while the slit
length is 7\farcm9. Details about the observations are presented in
Table~\ref{table1}, while the coordinates of the slit centers, the
number of spectra and their individual exposure times are given in
Table~\ref{table2}. The spectrophotometric standard stars HR5501,
HR7596, HR9087, HR718, and HR7950 (Hamuy et al. 1992, 1994) were
observed in order to calibrate the spectra.
\begin{table}
\caption[]{Spectral log}
\label{table2}
\begin{flushleft}
\begin{tabular}{lccc}
\noalign{\smallskip}
\hline
\hline
Area & \multicolumn{2}{c}{Slit centers} & No of spectra$^{\rm a}$ \\
 & R.A. & Dec. &  \\
\hline 
F2 &  21\h29\m18.5\s & 17\degr14\arcmin36\arcsec  & 1 \\
F3a &  21\h25\m15.2\s & 16\degr56\arcmin25\arcsec  & 2 \\
F3b(I) & 21\h23\m46.0\s & 17\degr28\arcmin38\arcsec  & 1\\
F3b(II) & 21\h23\m57.8\s & 17\degr23\arcmin35\arcsec  & 2 \\
F4 & 21\h22\m05.3\s & 17\degr25\arcmin35\arcsec  & 2 \\
F6(I) & 21\h14\m00.7\s & 20\degr02\arcmin34\arcsec  & 2 \\
F6(II) & 21\h14\m09.0\s & 20\degr05\arcmin12\arcsec  & 2 \\
\hline
\end{tabular}
\end{flushleft}
${\rm ^a}$ Single exposure time is 3600 sec\\
   \end{table}

\section{Imaging of newly detected structures}
\subsection{The \hnii\ mosaic}
Fig. 1 shows our \hnii\ mosaic. The image reveals considerable new
faint emission including filamentary and diffuse structures. Several
thin and curved filaments can be seen in this image and the most
interesting are described here in detail. Starting the morphological
study from the south--east area, an extremely faint thin filament
28\arcmin\ long is present (area F1 in Fig. 1 -- see also the
full resolution image in Fig. 2a), which lies a few arcminutes
to the south of the star HD 205300 with its center approximately at $\alpha
\simeq$ 21\h32\m50\s\ and $\delta \simeq$ 17\degr05\arcmin. A brighter
filamentary structure lies west of F1 (area F2 in Fig. 1, Fig. 2a),
where faint curved filaments and diffuse emission appear south of the
star HD 204560, extending down to a declination of
$\sim$17\degr04\arcmin\ ($\sim$40\arcmin\ long). The curvature of
these faint filaments is complex and it is difficult to define
their radii.  Three of the most prominent bright structures appear
to the south of the center of our mosaic between $\alpha
\simeq$ 21\h25\m50\s, $\delta \simeq$ 16\degr50\arcmin\ and $\alpha
\simeq$ 21\h21\m40\s, $\delta \simeq$ 17\degr26\arcmin\ (areas F3a, b
and F4 in Fig. 1, Fig. 3a). Arcs F3a, b are separated by a
$\sim$5\arcmin\ gap but their morphology suggests that they
belong to same filament. Diffuse emission is also present along the
$\sim$25\arcmin\ long arc F3a, which becomes brighter along arcs F3b
and F4. A large diffuse area covers a region of $\sim$1\degr\ north of
these filaments and exactly in the center of the \hnii\ mosaic.  Two
more new structures are present further to the west of this
diffuse emission. The first appears as a $\sim$1\degr\ long filament
(F5a, b in Fig. 1, Fig. 4a) separated by faint diffuse emission. The
second structure (F6 in Fig. 1, Fig. 5a) is located further to the
north at $\alpha \simeq$ 21\h13\m50\s\ and $\delta \simeq$
20\degr00\arcmin\ and seems to be correlated with a $\sim$1\degr\ long
faint structure. The latter, is north--west of F6 where faint
filamentary and diffuse emission is present.  Finally, there are two
more structures (areas F7, F8 in Fig. 1, Figs. 2b,c) which are located
at  $\alpha \simeq$ 21\h20\m40\s, $\delta \simeq$
20\degr07\arcmin very close to the multiple bright star HD 203504
(Tokovinin 1997) and $\alpha \simeq$ 21\h19\m50\s, $\delta
\simeq$ 21\degr40\arcmin, respectively. Both are dominated by diffuse
emission, while faint small scale filamentary structures are also
present. Emission also detected further to the north--west and
north--east of F8 seems to extend outside our field.

%--------------------------------------------------------
  \begin {figure*}
   \resizebox{\hsize}{!}{\includegraphics{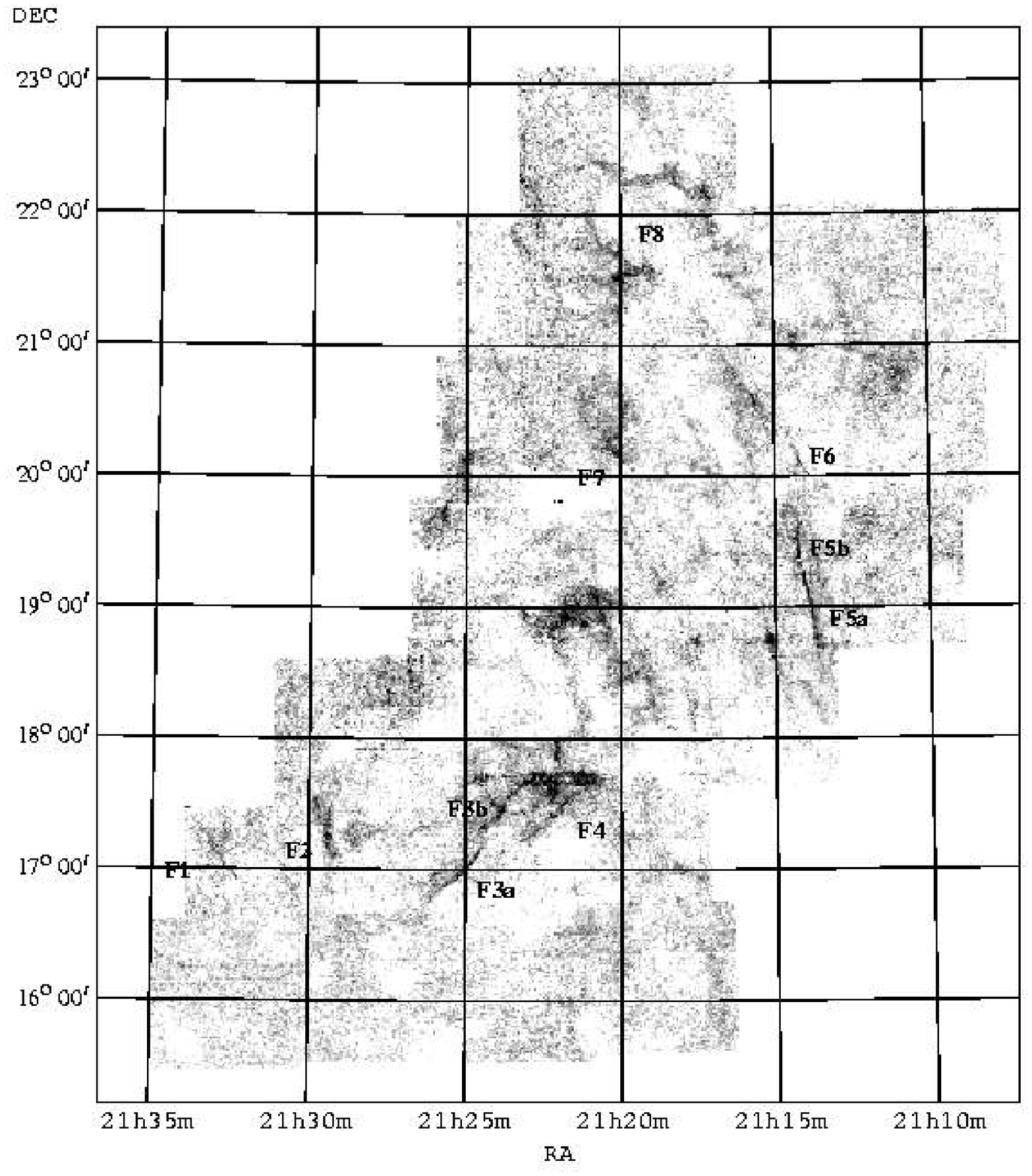}}
    \caption{The 7\degr.5 $\times$ 8\degr.5 mosaic showing the
    filamentary structures discovered in the low ionization lines of
    \hnii. Labels F1--F8 define the areas discussed in the text in
    more detail. The original 7500 $\times$ 8500 pixels$^{2}$~image
    was rebinned by a factor of 4 to produce this plot.}
     \label{fig01}
       \end{figure*}
%--------------------------------------------------------
  \begin {figure*}
   \resizebox{\hsize}{!}{\includegraphics{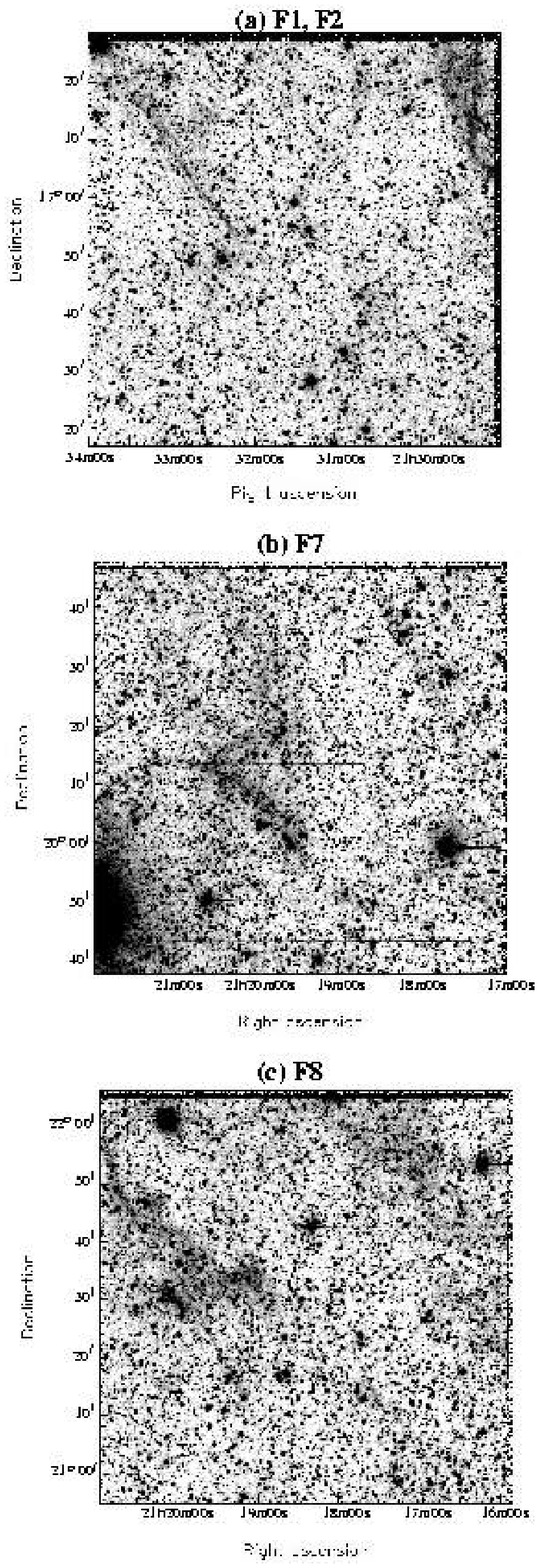}}
    \caption{Optical images in \hnii\ of filaments (a) F1, F2, (b)
    F7 and (c) F8.}
     \label{fig02}
  \end{figure*}
%--------------------------------------------------------
  \begin {figure*}
   \resizebox{\hsize}{!}{\includegraphics{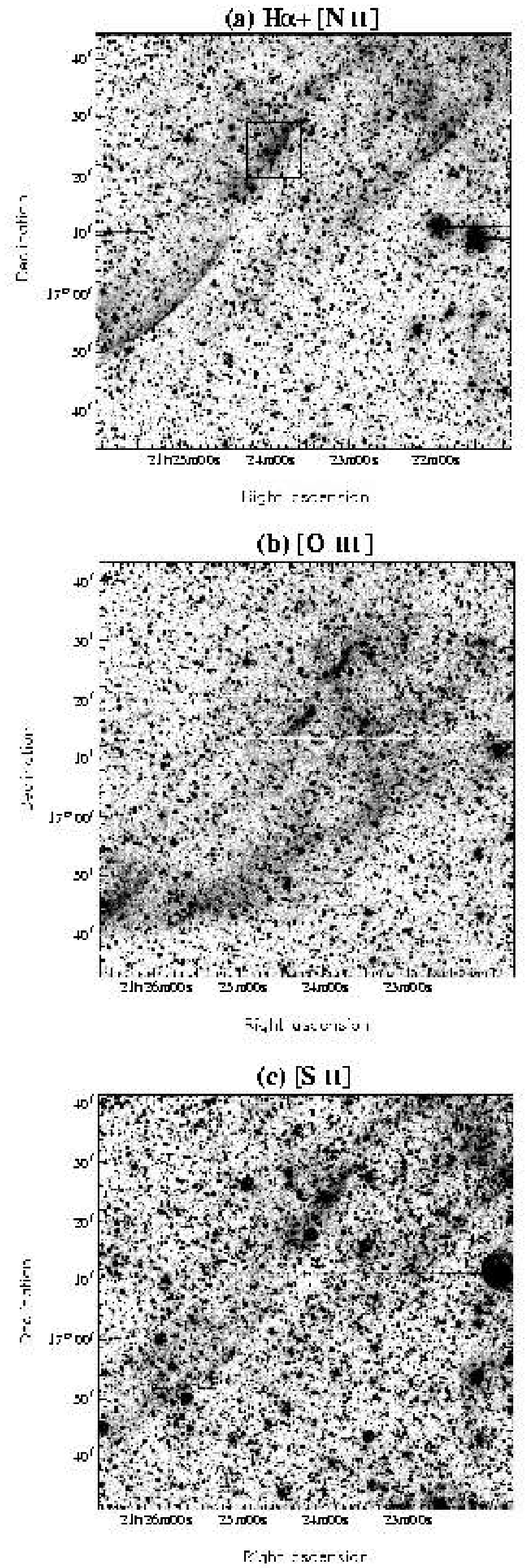}}
    \caption{Optical images of filaments F3 and F4 in (a) \hnii, (b)
    \oiii\ and (c) \sulfur. Note that the field in the \oiii\ and
    \sulfur\ images is shifted by $\sim$20\arcmin\ to the west because
    it was centered on filament F3. The rectangle in (a)
    represents the area which was imaged with the 1.3 m telescope and
    shown in Fig. 5a.}
     \label{fig03}
  \end{figure*}
%--------------------------------------------------------
  \begin {figure*}
   \resizebox{\hsize}{!}{\includegraphics{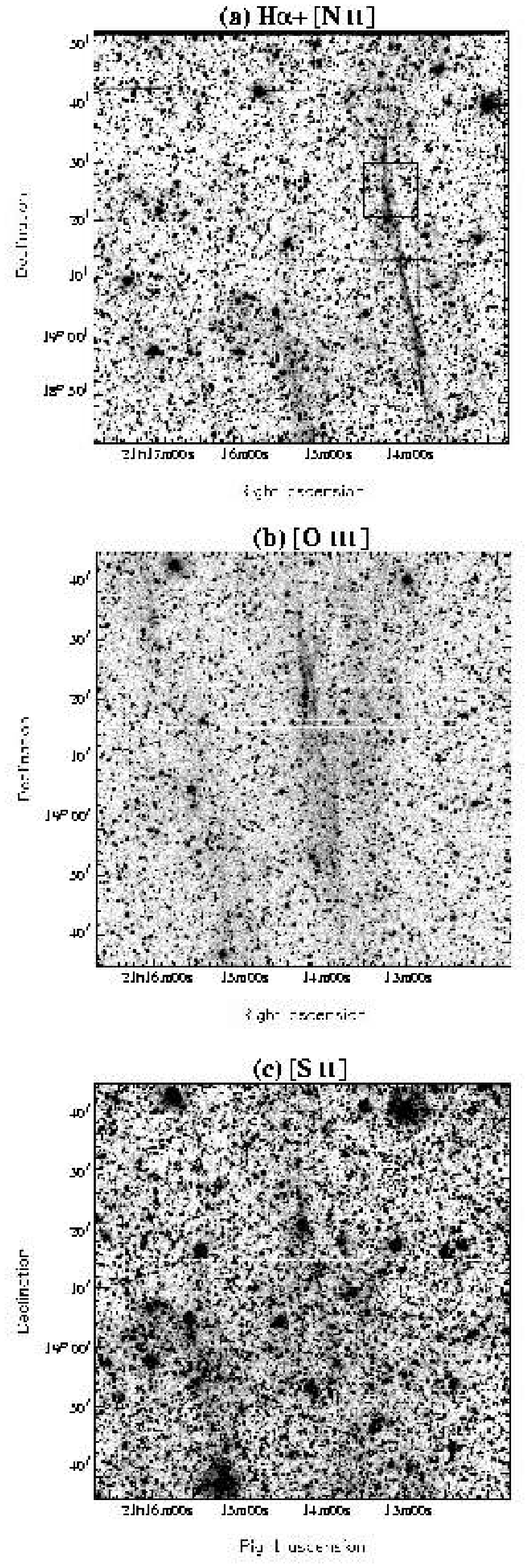}}
    \caption{Optical images of filament F5 in (a) \hnii, (b) \oiii\
    and (c) \sulfur. The field in the \oiii\ and \sulfur\ images is
    shifted by $\sim$15\arcmin\ east and $\sim$8\arcmin\ south because it
    was centered on filament F5. The rectangle in (a)
    represents the area imaged with the 1.3 m telescope and
    shown in Fig. 5b.}
     \label{fig04}
  \end{figure*}
%--------------------------------------------------------
  \begin {figure*}
   \resizebox{\hsize}{!}{\includegraphics{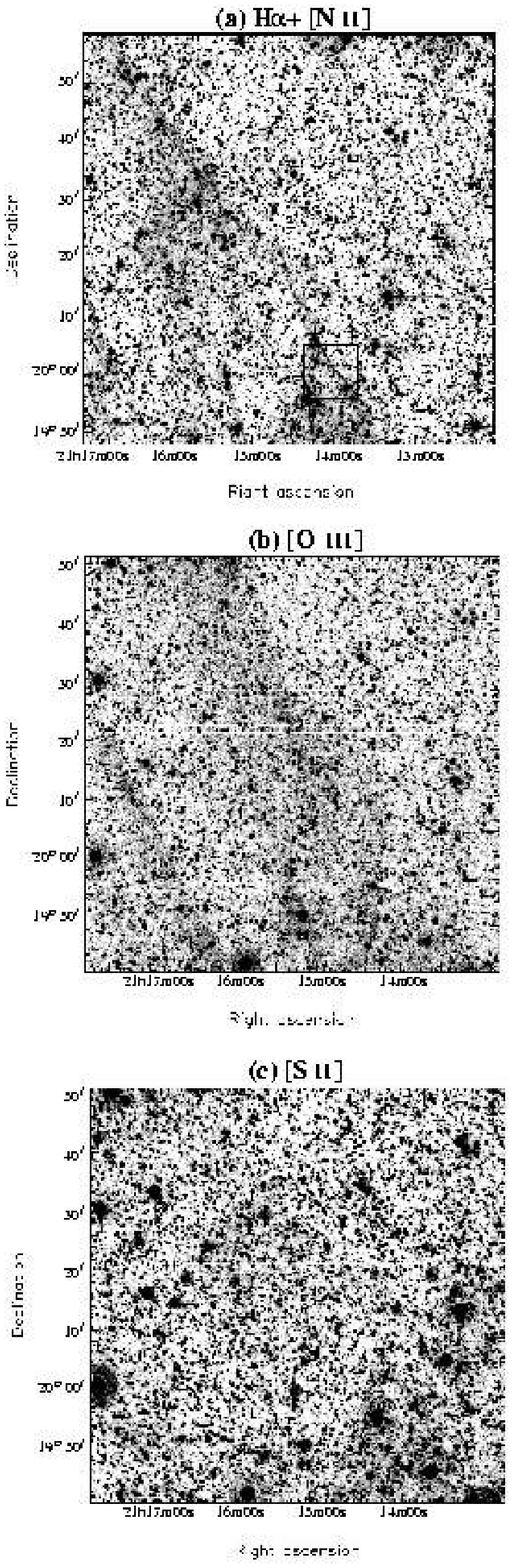}}
    \caption{Optical images of filament F6 in (a) \hnii, (b) \oiii\
    and (c) \sulfur. The field in the \oiii\ and \sulfur\ images is
    shifted by $\sim$11\arcmin\ west and $\sim$7\arcmin\ south because it
    was centered on filament F6. The rectangle in (a)
    represents the area observed with the 1.3 m telescope
    and shown in Fig. 5c.} 
     \label{fig05}
  \end{figure*}
\subsection{An intercomparison of the brightest filaments}
The \hnii\ images best describe the newly detected
structures. However, \sii\ and \oiii\ emission from structures F3--F6
is also detected and generally appears less filamentary and more
diffuse than in the \hnii\ image. Significant differences between the
\hnii, \oiii\ and \sii\ images are found for most of the filaments. In
filament F3 (Fig. \ref{fig03}), the arc--like structure F3b appears
intense and filamentary in the first two emission lines, while at the
same location the \sii\ emission is weaker and diffuse. Furthermore,
the filamentary structure F3a which is present in the \hnii\ image
does not have \oiii\ or \sii\ counterparts and only very faint diffuse
emission is found. On the other hand, small scale filamentary and
diffuse structures appear in all the emission line images of filament
F4 (Fig. \ref{fig03}), but still much better defined in \hnii\ than in
\oiii\ or \sii.  A similar situation appears in filament F5
(Fig. \ref{fig04}), where the long filamentary arc in \hnii\ is
separated by a $\sim$5\arcmin\ gap of faint diffuse emission. The
north part of this filament (F5b) appears both in \oiii\ and \sii,
while only diffuse emission is present at the location of filament
F5a. In the case of filament F6, the lower ionization lines of \hnii\
(Fig. \ref{fig05}a) display a different morphology than the medium
ionization line of \oiii\ (Fig. \ref{fig05}b). The small scale
filamentary structures which are present in the former do not appear
in the latter, where only diffuse emission is detected. Weak and
diffuse emission is also found in the low ionization image of \sii\
(Fig. \ref{fig05}c).

\subsection{The \hnii\ higher resolution images}
\begin {figure*}
   \resizebox{\hsize}{!}{\includegraphics{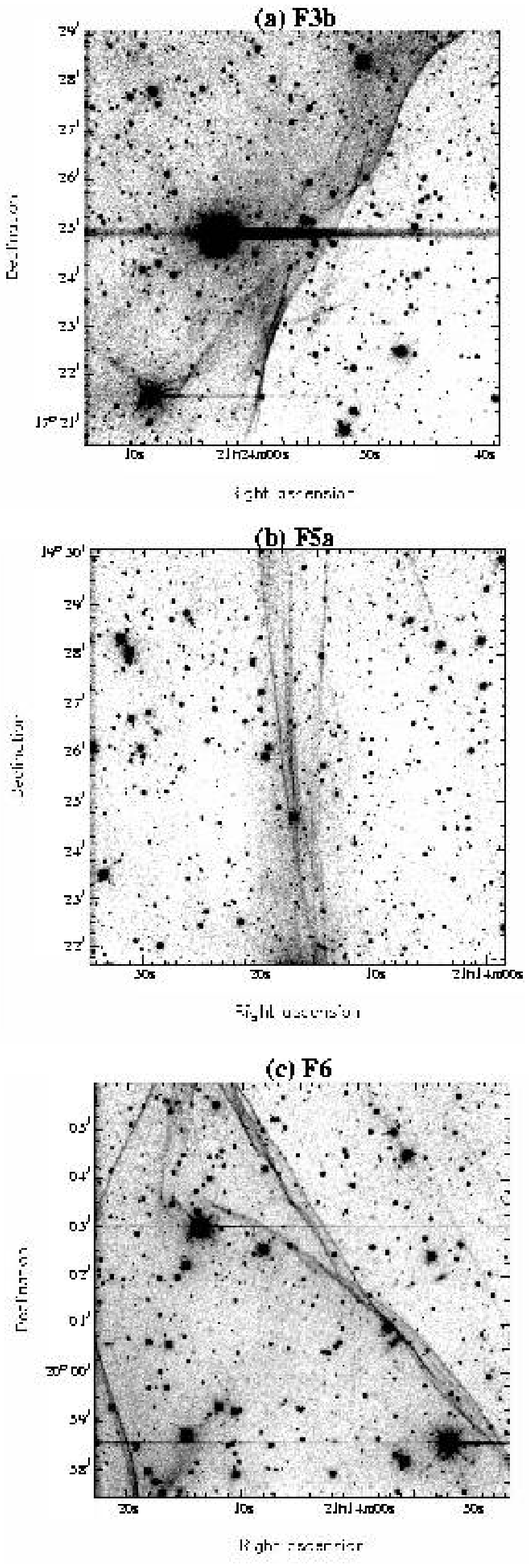}}
   \caption{Higher resolution greyscale representations (0\arcsec.5
   pixel$^{-1}$) of filaments (a) F3, (b) F5 and (c) F6, observed with
   the 1.3 m telescope in \hnii.}
     \label{fig06}
  \end{figure*}

Portions of selected filamentary structures (F3b, F5a and F6) have
been imaged in higher angular resolution (0\arcsec.5 pixel$^{-1}$)
with the 1.3 m telescope through the \hnii\ interference filter
(Fig. \ref{fig06}). All images show very thin filamentary structures
as well as diffuse emission. Starting from filament F3b (Fig. 6a) the
image shows a thin filament along the north--south direction separated
by a $\sim$1\arcmin\ gap in the center and extends outside the borders
of the image. This filament seems to separate the diffuse emission on
the left and the interstellar medium on the right. A thin and much
fainter complex of filamentary structures appears in the area of
filament F5a (Fig. 6b) which are almost parallel to the north--south
direction. In this area diffuse emission seems to be present only
close to these structures and mostly to the south. However, the
possibility that part of this diffuse emission may be stray light from
a bright star, just outside the south border of the image, cannot be
ruled out. Finally, in the area of filament F6 (Fig. 6c) a number of
new filamentary structures have been discovered to the east and west,
which cover the north--south direction across the image and extending
outside the borders of the 8\arcmin.5 $\times$ 8\arcmin.5 square
area. The filament F6 located in the right side of the image
consists of two bright and two faint thin filaments. In addition, we
detected one bright filament in the lower left side of the image and
three faint filaments in the north. Diffuse emission is again present
to the east part of this area. Note that the exact position and the
area covered by the higher resolution images are marked in Figs. 3a,
4a and 5a with solid rectangles.

%--------------------------------------------------------
\section{The long--slit spectra}
The deep low resolution spectra were taken on the relatively bright
optical filaments at seven different locations (their exact positions
are given in Table~\ref{table2}). In Table~\ref{sfluxes}, we present
the relative line fluxes taken from different apertures along the
slit, that are free of field stars and include sufficient
line emission to allow an accurate determination of the observed
lines. The background extraction apertures were selected towards the
north and south ends of each slit depending on the filament's position
within the slit. The signal to noise ratios presented in
Table~\ref{sfluxes}~do not include calibration errors which are
typically less than 10\%. The absolute \ha\ flux covers a range of
values from 1.1 to 8.8 $\times$ \flux. The extracted apertures do not
show the same nature of emission. There are areas (e.g. filament
F3b(IIb), F6(Ia) etc.) which show that the observed optical emission
must originate from shock heated gas, since \sii/\ha\ $>$0.4 (Smith et
al. 1993), while other areas (e.g. F3a(Ia), F3b(I) etc.) display
\sii/\ha\ $<$0.3 and therefore suggest photoionization processes.
\par
The \siirat\ ratio of $>$1.1 indicates low electron densities (less
than 440 cm$^{-3}$; Osterbrock 1989). However, taking into
account the statistical errors on the sulfur lines, we calculate with
the nebular package (Shaw \& Dufour 1995), that electron densities up
to 600 cm$^{-3}$ are allowed.  The logarithmic extinction c
towards a source of line emission was calculated by using the
\ha/\hbeta\ observed ratios and an intrinsic ratio of 2.85 (see
Sect. 5 for details). Furthermore, \oiii\ emission was only
detected in filaments F3b and F6(II), while in the rest of the
filaments only upper limits are given. The \oiii/\hbeta\ ratio was
measured and in almost all cases was found to be less than 4. The
theoretical models of Cox \& Raymond (1985) and Hartigan et al. (1987)
suggest that shocks with complete recombination zones must be present,
while shock velocities  $\leq$100 \vel\ are estimated.

\begin{table*}
\caption[]{Relative line fluxes.}
\label{sfluxes}         
\begin{tabular}{lllllllll}
\hline
\hline  
\noalign{\smallskip}
Line (\AA) & F2(Ia) & F2(Ib) & F3a(Ia)$^{\rm a}$ & F3a(Ib)$^{\rm a}$ & F3b(I) & F3b(IIa)$^{\rm a}$ & F3b(IIb)$^{\rm a}$ & F4$^{\rm a}$  \\
\hline
4861 \hbeta\ & $<$ 12 & $<$ 16 & 19 (9) & 15 (3) & 23 (5) & 28 (20) & $<$ 17 & 16 (4) \\
5007 \oxygen\ & $<$ 9 & $<$ 15 & $<$ 10 & $<$ 15 & 33 (8) & $<$ 6 & 55 (25) & $<$ 15  \\
6548 \nii\  & 6 (4) & $-$ & $-$ & $-$ & 6 (3) & $-$ & 28 (22) & 9 (5) \\
6563 \ha\ & 100 (62) & 100 (52) & 100 (89) & 100 (66) & 100 (41) & 100 (142) & 100 (74) & 100 (53)  \\
6584 \nii\  & 27 (19) & 25 (14) & $-$ & $-$ & 18 (8) & 21 (34) & 93 (71) & 23 (13) \\
6716 \sii\ & 18 (16) & 21 (11) & 8 (9) & 19 (17) & 12 (6) & 12 (22) & 63 (54) & 24 (16) \\
6731 \sii\  & 13 (12) & 20 (10) & 4 (5) & 17 (16) & 7 (4) & 6 (11) & 45 (40) & 21 (15) \\
\hline
Absolute \ha\ flux$^{\rm b}$ & 3.7 & 3.0 & 6.7 & 3.1 & 3.1 & 8.8 & 3.1 & 4.1 \\
\ha /\hbeta\   & $-$ & $-$ & 5.2 (9) & 6.7 (3) & 4.4 (5) & 3.6 (20) & $-$ &  6.1 (4) \\
\sii/\ha\ & 0.3 (19) & 0.4 (14) & 0.1 (10) & 0.4 (22) &  0.2 (7) & 0.2 (23) & 1.1 (50) & 0.5 (20) \\
F(6716)/F(6731)	& 1.4 (9) & 1.1 (7) & 1.9 (4) & 1.1 (12) & 1.6 (3) & 2.1 (10) & 1.4 (32) & 1.1 (11) \\
\oxygen/\hbeta\ & $-$ & $-$ & $<$ 1.5 & $<$ 1.0 & 1.4 & $<$ 0.2 & $>$ 3.2 & $<$ 0.9 \\
c & $-$ & $-$ & 0.77 (6) & 1.06 (3) & 0.53 (3) & 0.28 (4) & $-$ & 0.98 (3) \\
\hline 
\end{tabular}

\begin{tabular}{llll}
\hline
 \noalign{\smallskip}
Line (\AA) & F6(Ia)$^{\rm a}$ & F6(Ib)$^{\rm a}$ & F6(II)$^{\rm a}$ \\
\hline
4861 \hbeta\ & 25 (3) & 18 (8)  & $<$ 6  \\
5007 \oxygen\ & $<$ 27 & $<$ 7 & 10 (11)  \\
6548 \nii\   & 10 (2) & 5 (3) & $-$ \\
6563 \ha\    & 100 (26) & 100 (70) & 100 (142)  \\
6584 \nii\   & 48 (13) & 11 (9) & 11 (7)  \\
6716 \sii\   & 32 (10) & 7 (6) & 14 (26)  \\
6731 \sii\   & 24 (8) & 4 (3) & 4 (8)  \\
\hline
Absolute \ha\ flux & 1.1 & 6.9 & 6.7 \\
\ha /\hbeta\  & 4.0 (3) & 5.7 (7) & $-$  \\
\sii/\ha\     & 0.6 (11) & 0.1 (7) & 0.2 (24)  \\
F(6716)/F(6731)	& 1.3 (6) & 1.9 (3) & 3.1 (8)  \\
\oxygen/\hbeta\ & $<$ 1.1 & $<$ 0.4 & $>$ 1.7 \\
c & 0.42 (3) & 0.83 (5) & $-$\\
\hline 
\end{tabular}

${\rm ^a}$ Listed fluxes are a signal to noise weighted
average of two fluxes
 
$^{\rm b}$ In units of \flux\

All fluxes normalized to
F(H$\alpha$)=100 and are uncorrected for interstellar extinction. Numbers in
parentheses represent the signal to noise ratio of the quoted fluxes.\\
\end{table*}
\section{The ROSAT All--Sky Survey data}
The area in the Pegasus Constellation was observed by ROSAT during the
All--Sky Survey for $\sim$500 s. Very weak and diffuse emission is
present in the area where optical line emission is detected
(Fig. 7). The typical level of X--ray emission in the central areas of
our field is $\sim 1.5 \times 10^{-3}$~cts s$^{-1}$~arcmin$^{-2}$.
Note that a lane of very low X--ray emission to the east has a much
more lower exposure time ($\sim$120 s). Unfortunately, the faintness
of the soft X--ray emission did not allow us to obtain reliable
results from the spectral analysis.

%--------------------------------------------------------
  \begin {figure*}
   \resizebox{\hsize}{!}{\includegraphics{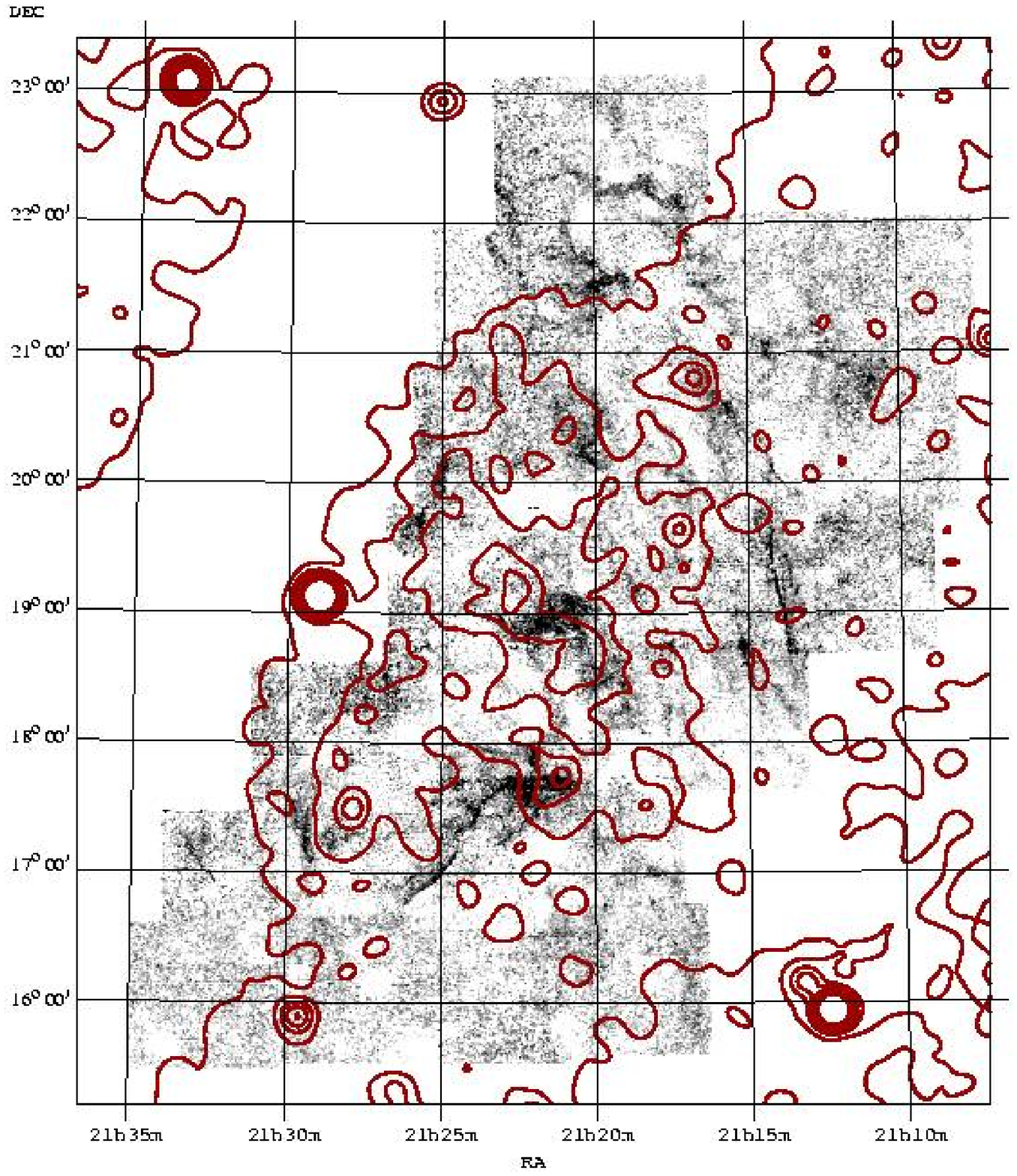}}
   \caption{The correlation between the soft X--ray emission
   (contours) and the \hnii\ emission of the field. The X--ray
   contours scale linearly from 9.6$\times 10^{-4}$~cts s$^{-1}$
   pixel$^{-1}$~to 3.6$\times 10^{-3}$~cts s$^{-1}$ pixel$^{-1}$,
   while each pixel is 1\arcmin.0$\times$1\arcmin.0.}
     \label{fig07}
  \end{figure*}
%--------------------------------------------------------
%
\section{Discussion}
The newly discovered filamentary structures towards the Pegasus
constellation show up as incomplete circular structures in the
optical, as diffuse emission in the soft X--ray wavelengths
and without any radio emission detected so far. The low ionization
\hnii\ mosaic shows several filaments across the observed field, while
the selected \sii\ and
\oiii\ images generally display diffuse emission. The weak soft X--ray
emission seems to overlap the optical emission. The long--slit spectra
suggest that part of the detected emission results from shock heated
gas, while in other locations the emission seems to result from
photoionization processes. Low \sii/\ha\ ratios are indicative of
\HII\ regions, while higher values support the interpretation of an
expanding SNR shell.
\par
The interstellar reddening was derived from
the \ha/\hbeta\ ratio (Osterbrock 1989), using the interstellar extinction
law by Fitzpatrick (1999) and R$_{\rm v} = 3.1$~for all
objects. Therefore, the interstellar extinction c(\hbeta) can be
derived from the relationship

\begin{equation}
{\rm c(H\beta)} = \frac{1}{0.348} \log\frac{{\rm F(H\alpha)}/{\rm
F(H\beta)}}{2.85},
\end{equation} 

where 0.348 is the relative logarithmic extinction coefficient for
\hbeta/\ha\ and 2.85 the theoretical value of F(\ha)/F(\hbeta).
Consequently, the interstellar extinction c at the positions of
different filaments (see Table~\ref{sfluxes}) was found between
$\sim$0.3 and 1.0 or an A$_{\rm V}$~of $\sim$0.6 to 2.1. In Table
\ref{sfluxes}, we have also calculated the estimated signal to noise for each
extinction value. We propose that an interstellar extinction c of
$\sim$0.7--0.8 or an A$_{\rm V}$~of $\sim$1.4--1.6 is representative
of the extinction towards this area.
\par
An estimated value of the hydrogen column density N$_{\rm H}~{\rm of}
\sim 1 \times 10^{21}$~cm$^{-2}$ is given by Dickey \& Lockman (1990)
in the direction of the optical filaments. Using the statistical
relation of Predehl \& Schmitt (1995)

\begin{equation}
{\rm N_{H}} = 5.4~(\pm 0.1) \times 10^{21}~{\rm E_{B-V}}~{\rm cm}^{-2},
\end{equation}
 
where A$_{{\rm V}} = 3.1 \times {\rm E_{B-V}}$~(Kaler 1976), we obtain
an N$_{{\rm H}}$~of $1.0 \times 10^{21}~{\rm cm}^{-2}$~and $3.8 \times
10^{21}~{\rm cm}^{-2}$~for the minimum and maximum c values calculated
from our spectra, respectively. These values are consistent with the
estimated galactic N$_{{\rm H}}$~considering the uncertainties
involved.
\par
The electron densities have been
determined by measuring the ratio of \siirat\ and found densities
below $\sim$440 \dens. In case of complete recombination zones, it is
possible to estimate the preshock cloud density n$_{\rm c}$ by using
the relationship (Dopita 1979)

\begin{equation}
{\rm n_{[SII]} \simeq\ 45\ n_c V_{\rm s}^2}~{\rm cm^{-3}},
\end{equation}

where ${\rm n_{[SII]}}$ is the electron density derived from the
sulfur line ratio and V$_{\rm s}$ is the shock velocity into the
clouds in units of 100 \vel . Thus, we obtain ${\rm n_c} V_{\rm s}^2
< 10$~from Eq. (3). Shock velocities below 100 \vel\ are consistent
with our data suggesting that the preshock cloud densities are
typically around or below 10 \dens.
\par
Both optical and X--ray observations do not uniquely identify the
nature of the detected emission. The optical observations suggest the
existence of
\HII\ regions in the area as well as shock--heated filamentary
structures. However, the presence of soft X--ray emission which seems
to overlap the optical emission may not be a chance coincidence and
would provide support to the scenario where the optical filaments may
be part of an extended supernova remnant. Shock--heated emission was
found in almost all the filamentary structures and taking into account
their morphology it may be possible that they belong to a single
extended SNR with its east part lying outside our field. 
The overall orientation of the filaments suggests an estimated center
of the possible supernova event at around $\alpha \simeq$ 21\h24\m\
and $\delta \simeq$ 19\degr30\arcmin\  (equivalent to galactic longitude $l
\simeq 70^{o}.1$ and latitude $b \simeq -21^{o}.5$) with a
angular diameter of $\sim$4\degr.5$\times$6\degr. Large--scale
H$\alpha$ surveys also show evidence of optical emission in the area
and its morphology gives support to the SNR scenario. In particular,
the Wisconsin H$\alpha$~Mapper (WHAM -- Haffner et al. 2002) shows a
very faint elliptical distribution of emission almost at the same
location with its center at $l \simeq 69^{o}.8$, $b \simeq -20^{o}.8$
and an angular size of $\sim$4\degr$\times$6\degr, while several maps
of the Virginia Tech Spectral-Line Survey (VTSS -- Dennison et
al. 1998) suggest a roughly spherical distribution in the area of the
brighter emission. Duncan et al. (1997) presented a number of large
($>$ 2\degr) SNR candidates where almost all appear as
edge--brightened and filamentary arcs, detected towards high latitudes
away from the Galactic plane. They concluded that these structures
must be large and old, within the radiative phase and almost at the
point of merging with the interstellar medium (ISM). If our newly
detected structures form indeed a SNR, then it is probably in the late
stages of evolution where it begins to merge to the ISM. This may
explain why no radio emission has been detected from radio surveys
since as it becomes older it is difficult to distinguish its emission
from the radio emission of the galactic plane (Duncan et
al. 1997). Assuming a distance to the SNR of 1 kpc and an angular
diameter of $\sim$5\degr\ then its radius would be $\sim$44 pc. If
this is the case then this SNR would be one of the largest known
SNRs. However, it must be pointed out that the possibility of the
existence of more than one supernova explosions cannot be excluded
since the unclear nature of both optical and X--rays does not help to
clarify which scenario is more probable.
\par
The observed variation in the absolute line fluxes could be due to
variations of the interstellar clouds density, shock velocities,
intrinsic absorption, while the absence of the X--ray emission in the
north--west area suggests even lower interstellar medium densities in
that region.  According to Chu (1997), when a supernova event occurs
in a hot, low--density medium, the classical SNR signatures may not be
observable and mostly X--ray diffuse emission is present. Massive
stars are usually formed in groups, such as OB associations, thus the
possibility that the optical emission is associated to a supernova
remnant within an OB association must be examined further in the
future. If the optical emission belongs to a supernova remnant, then
it would be one of the largest remnants and this may explain why,
unlike other known SNR, it is not observed as a source of non--thermal
emission. The origin of these large scale structure is not yet
understood especially at this high galactic latitude and deeper
observation in radio and X--ray wavelengths would be required to
identify its nature.
\par
Three pulsars (PSRs
2113$+$2754, 2116$+$1414 and 2124$+$1407) have been found within a few
degrees away from the brightest filaments (Taylor et al. 1995). These
pulsars rotate at a period of 601, 440 and 694 ms, at estimated
distances of 1.37, 2.11 and 2.08 kpc, respectively. An estimate of the
proper motion of PSRs 2113$+$2754 and 2116$+$1414 was determined by
Harrison et al. (1993) who measured $\mu_{\alpha}=-23~{\rm mas\
yr}^{-1}$, $\mu_{\delta}=-54~{\rm mas\ yr}^{-1}$ and
$\mu_{\alpha}=8~{\rm mas\ yr}^{-1}$, $\mu_{\delta}=-11~{\rm mas\
yr}^{-1}$, respectively.  It is not clear at the moment if any of
these pulsars is related to the supernova event(s) associated to the
optical filaments but a correlation can not be ruled out.
\section{Conclusions}
Unknown filamentary and diffuse structures have been discovered in
Pegasus constellation through deep imaging and spectral
observations. These structures appear more filamentary in \hnii\
emission lines images than the \oiii\ and \sii\ images. Diffuse
X--ray emission is detected in the area and may be associated with the
optical line emission. The flux calibrated long--slit spectra
indicate that the emission arises from both photoionized and shock
heated gas. The \siirat\ ratio indicates low electron densities while
shock velocities around 100 \vel\ were found. The equivalent hydrogen
column density was estimated from the extinction derived from
\ha/\hbeta\ ratio to be between 1.0 to 3.8 $\times
10^{21}$~cm$^{-2}$. It may be possible that most of the optical
emission is part of one or more supernova remnants.
\begin{acknowledgements}
We would like to thank the referee R. G. Arendt for his comments which
helped to clarify the scope of this paper. Skinakas Observatory is a
collaborative project of the University of Crete, the Foundation for
Research and Technology-Hellas and the Max-Planck-Institut f\"ur
Extraterrestrische Physik. This research has made use of data
obtained through the High Energy Astrophysics Science Archive Research
Center Online Service, provided by the NASA/Goddard Space Flight
Center. The Wisconsin H-Alpha Mapper (WHAM) and the Virginia Tech
Spectral-Line Survey (VTSS) are funded by the National Science
Foundation.
\end{acknowledgements}
%

%---------------
%

%

\begin{thebibliography}{}

        \bibitem[1989]{are89} Arendt R. G., 1989, ApJS, 70, 181

        \bibitem[2001]{bou01} Boumis P., Dickinson C., Meaburn J., Goudis 
                C.D., Christopoulou P. E., Lopez J. A., Bryce M., Redman 
                M. P., 2001, MNRAS, 320, 61 

        \bibitem[2002]{bou02} Boumis P., Mavromatakis F. and
                Paleologou, E. V., 2002, A\&A, 385, 1042		 

        \bibitem[1997]{chu97} Chu Y-H., 1997, AJ, 113, 1815

	\bibitem[1985]{cox85} Cox D. P. and Raymond J. C., 1985, 
		ApJ, 298, 651

        \bibitem[1998]{den98} Dennison B., Simonetti J. H. \& Topasna G.,
                1998, PASA, 15, 147 (The Virginia Tech Spectral--Line Survey,
                http://www.phys.vt.edu/~halpha/)

	\bibitem[1990]{dic90} Dickey J. M. and Lockman F. J., 1990, ARAA,
                28, 215

	\bibitem[1979]{dop79} Dopita M. A., 1979, ApJS, 40, 455

        \bibitem[1997]{dun97} Duncan A. R., Stewart R. T., Haynes R. F. and
                Jones K. L., 1997, MNRAS, 287, 722

	\bibitem[1995]{fes95} Fesen R. A. and Hurford A. P., 1995, AJ, 110, 
		747

	\bibitem[1997]{fes97} Fesen R. A., Downes R. A., Wallace D. and
                Normandeau M., 1995, AJ, 110, 2876

	\bibitem[1997]{fes97} Fesen R. A., Winkler P. F., Rathore Y., 
		Downes R. A., Wallace D. and Tweedy R. W., 1997, AJ, 113, 767

	\bibitem[1999]{fit99} Fitzpatrick E. L., 1999, PASP, 111, 63

        \bibitem[2001]{gre01} Green D. A., 2001, A Catalog of Galactic
                Supernova Remnants (2001 version), Mullard Radio Astronomy
                Observatory, Cavendish Laboratory, Cambridge, UK 
                (http://www.mrao.cam.ac.uk/surveys/snrs/)

         \bibitem[2002]{haf02} Haffner, L. M., Reynolds, R. J., Tufte, S.
                 L., et al., 2002, The Wisconsin H$\alpha$~Mapper, in prep 
                 (http://www.astro.wisc.edu/wham/)

         \bibitem[1992]{ham92} Hamuy M., Walker A. R., Suntzeff N. B.,
                 Gigoux P., Heathcote S. R. \& Phillips M. M., 1992, PASP, 
                 104, 533

         \bibitem[1994]{ham94} Hamuy M., Suntzeff N. B., Heathcote S. R.,
                 Walker A. R., Gigoux P. \& Phillips M. M., 1994, PASP, 106,
                 566

         \bibitem[1993]{har93} Harrison P. A., Lyne A. G. \& Anderson B., 1993,
                 MNRAS, 261, 113

        \bibitem[1987]{har87} Hartigan P., Raymond J. and Hartmann L., 1987,
                 ApJ, 316, 323

	\bibitem[1976]{kal76} Kaler J. B., 1976, ApJS, 31, 517

        \bibitem[1991]{koo91} Koo B.-C. and Heiles C., 1991, ApJ, 382, 204

        \bibitem[1999]{Las99} Lasker B. M., Russel J. N. \& Jenkner H., 1999,
                in the HST Guide Star Catalog, version 1.1-ACT, The 
                Association of Universities for Research in Astronomy, Inc

        \bibitem[2001]{mav01} Mavromatakis, F., Papamastorakis, J.,
                Ventura, J., Becker, W., Paleologou, E. V. and Schaudel, 2001,
                A\&A 370, 265

        \bibitem[2002]{mav02} Mavromatakis, F., and Strom R. G., 2002, A\&A,
                382, 291

        \bibitem[2002]{mav02a} Mavromatakis, F., Boumis P. and
                Paleologou, E. V., 2002a, A\&A, 383, 1011

        \bibitem[2002]{mav02c} Mavromatakis, F., Boumis P. and
                Paleologou, E. V., 2002b, A\&A, 387, 635

        \bibitem[2002]{mav02b} Mavromatakis, F., Boumis P., Papamastorakis J.
                and Ventura J., 2002c, A\&A, 388, 355

        \bibitem[1989]{ost89} Osterbrock D. E., 1989, Astrophysics of
                gaseous nebulae, W. H. Freeman \& Company

        \bibitem[1995]{pre95} Predehl P. and Schmitt J. H. M. M., 1995, A\&A, 
                 293, 889

        \bibitem[1992]{sak92} Saken J. M., Fesen R. A. and Shull J. M., 1992,
                ApJS, 81, 715

        \bibitem[1990]{sew90} Seward F. D., 1990, ApJS, 73, 781

        \bibitem[1995]{sha95} Shaw R. A. and Dufour R. J., 1995, PASP, 107, 896

        \bibitem[1993]{smi93} Smith R. C., Kirshner R. P., Blair W. P.,
                Long K. S. and Winkler P. F., 1993, ApJ, 407, 564  

        \bibitem[1995]{tay95} Taylor J. H., Manchester R. N., Lyne A. G., and
                Camilo F., 1995, A Catalog of 706 Pulsars, Princeton Pulsar 
                Group, Princeton University

        \bibitem[1997]{tok97} Tokovinin A. A., 1997, A\&AS, 124, 75

%
%
\end{thebibliography}
\end{document}